\documentclass[%
superscriptaddress,
reprint,
 amsmath,amssymb,
 aps,
 prb,
 ]{revtex4-2}

\usepackage{amsmath,amssymb}
\usepackage{amscd, latexsym}
\usepackage{mathrsfs}
\usepackage{graphicx}
\usepackage{epstopdf}
\usepackage{cancel}
\usepackage{amsfonts}
\usepackage{exscale}
\usepackage{dcolumn}
\usepackage{bm}
\usepackage{color}
\usepackage{xcolor}
 \usepackage{natbib}
\usepackage{lmodern}
\usepackage[T1]{fontenc}
\usepackage{verbatim}
\usepackage{float}
\usepackage {hyperref}
\hypersetup{colorlinks=true}

\usepackage{siunitx}

\usepackage[normalem]{ulem}

\newcommand{\ket}[1]{\left| #1 \right>} 
\newcommand{\bra}[1]{\left\langle #1 \right|}

\makeatletter
\newsavebox{\@brx}
\newcommand{\llangle}[1][]{\savebox{\@brx}{\(\m@th{#1\langle}\)}%
  \mathopen{\copy\@brx\kern-0.5\wd\@brx\usebox{\@brx}}}
\newcommand{\rrangle}[1][]{\savebox{\@brx}{\(\m@th{#1\rangle}\)}%
  \mathclose{\copy\@brx\kern-0.5\wd\@brx\usebox{\@brx}}}
\makeatother

\begin{document}

\title{Excitation dynamics in inductively coupled fluxonium circuits}

\author{A. Bar\i\c{s} \"Ozg\"uler}
\altaffiliation{The initial part of the work was performed at the University of Wisconsin--Madison.}
\affiliation{%
  Fermi National Accelerator Laboratory, Batavia, Illinois, 60510
}%
\author{Vladimir E. Manucharyan}
\affiliation{%
  Department of Physics, Joint Quantum Institute, and Center for Nanophysics and Advanced Materials, University of Maryland, College Park, MD 20742
}%
\author{Maxim G. Vavilov}
\affiliation{%
Department of Physics, University of Wisconsin--Madison, Madison, WI 53706
}%

\date{April 7, 2021}

\begin{abstract}

We propose a  near-term quantum simulator based on the fluxonium qubits inductively coupled to form a chain. This system provides long coherence time, large anharmonicity, and strong coupling, making it suitable to study Ising spin models. At the half-flux quantum sweet spot, the system is described by the transverse field Ising model (TFIM). 
We evaluate the propagation of qubit excitations through the system. 
As disorder increases, the excitations become localized. A single qubit measurement using the circuit QED methods is sufficient to identify localization transition without introducing tunable couplers. 
We argue that inductively coupled fluxoniums provide opportunities to study localization and many-body effects in highly coherent quantum systems.

\end{abstract}

\maketitle

\emph{Introduction --} Classical simulators hit their limitations of analyzing large systems with many degrees of freedom as the Hilbert space grows exponentially with the system size \cite{zhou2020limits}. Quantum simulators have potential to push the limits \cite{alexeev2021quantum, altman2021quantum}.
Digital quantum simulators (DQS)  use gates to approximate the unitary evolution, whereas analog quantum simulators (AQS) mimic the time evolution given by the Hamiltonian of another system by its controllable and tunable components \cite{georgescu2014quantum}. AQS can currently simulate a small class of Hamiltonian models, but they are not as prone to the Trotterization and gate errors as DQS, so AQS are practical for the NISQ era \cite{preskill2018quantum}. 

Quantum simulators have been performed in many platforms such as atomic spins \cite{edwards2010quantum}, vacancy centers~\cite{lei2017decoherence}, Rydberg atoms \cite{bernien2017probing}, trapped ions \cite{smith2016many}, ultracold atoms \cite{bordia2017periodically}, optical lattices \cite{schreiber2015observation} and superconducting qubits \cite{arute2020observation}. 
Transmons have been one of the most successful and widely used superconducting qubits due to the ease of their engineering and decent coherence properties \cite{las2014digital, leib2016transmon, tacchino2019digital, arute2019quantum, krantz2019quantum, kjaergaard2020superconducting}. The effect of charge noise is reduced in transmons due to capacitative shunting, but transmons are far from perfect two-level systems due to their small anharmonicity. An alternative qubit that is still protected from the low-frequency charge noise but has no sacrifice in anharmonicity is the fluxonium qubit \cite{manucharyan2009fluxonium,nguyen2019high, catelani2019fluxonium, somoroff2021millisecond}.
In addition to its long coherence, the fluxonium can be designed to have multiple strong connections with its neighbors via galvanic coupling~\cite{kou2017fluxonium}.

\begin{figure}[!htbp]
\begin{centering}
\includegraphics[width=0.9\columnwidth]{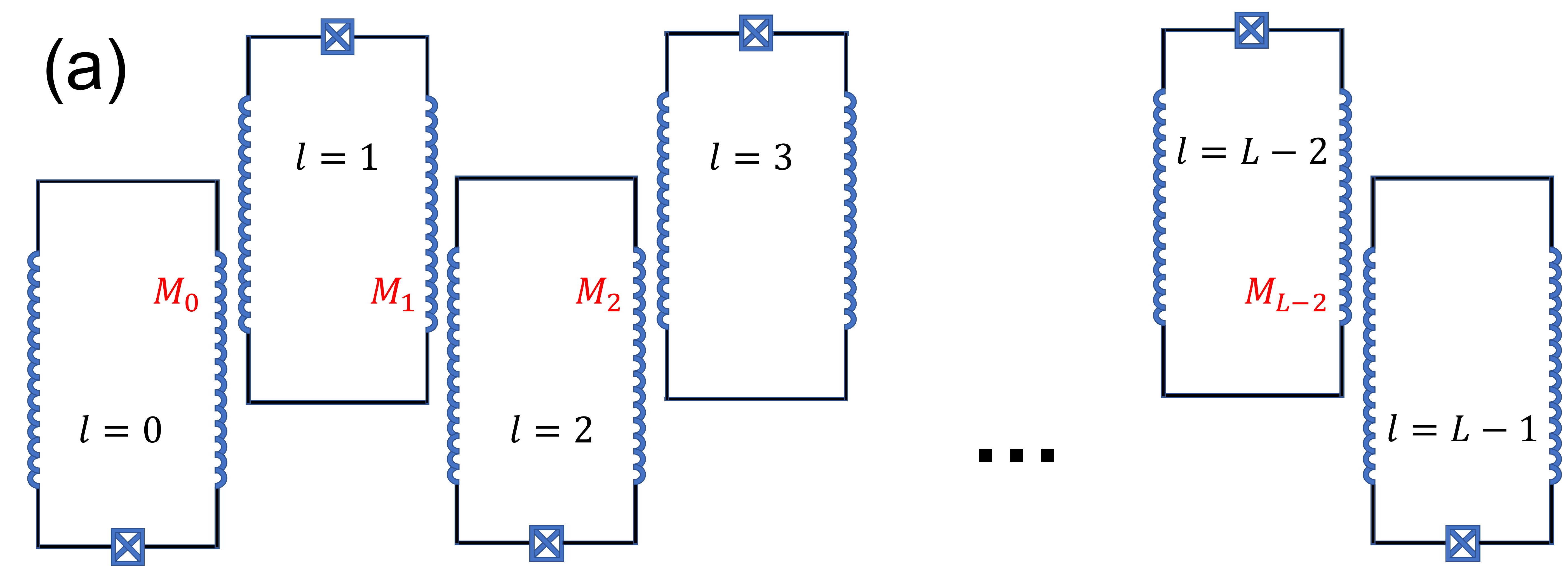}
\includegraphics[width =0.7\columnwidth]{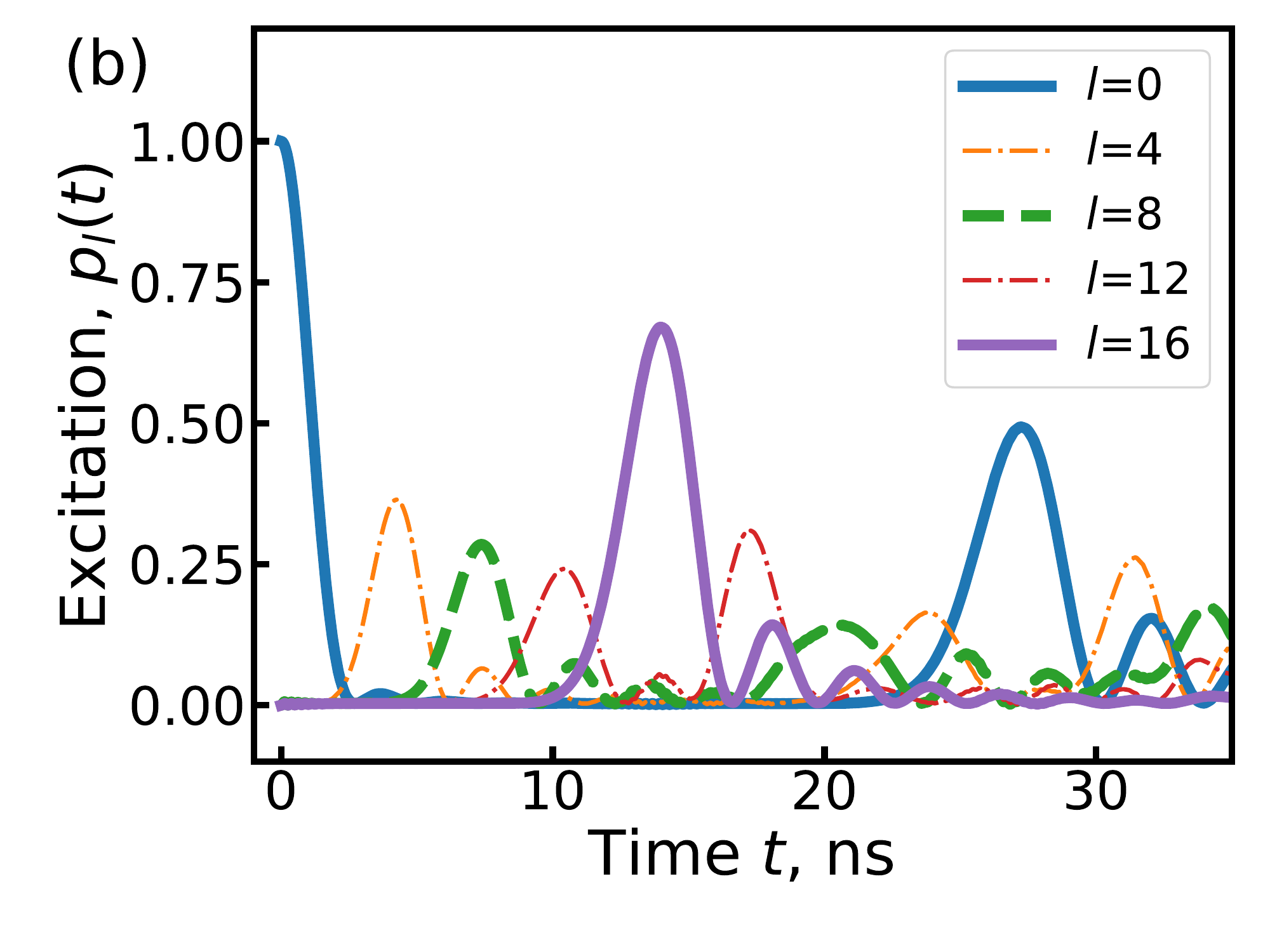}
\par\end{centering}
\caption{\label{fig:FluxoniumCircuit_and_EnergyDiagram}
(a) Open chain of $L$ fluxonium qubits with nearest neighbor inductive coupling.
(b) Time dependence of the excitation probability for qubits $l=0,4,8,12,16$ in the clean system of $L=17$ fluxoniums. Initially, all qubits were in their ground states, but $l=0$ was excited.  The parameters of fluxoniums and the coupling strength are provided in the text.
}
\label{fig:1}
\end{figure}

In this paper, we demonstrate that a fluxonium chain (Fig.~\ref{fig:1}(a)) is an efficient tool for simulations of clean and disordered transverse field Ising model (TFIM)~\cite{suzuki2012quantum}.  We study the dynamics of excitation propagation through a one-dimensional chain of inductively coupled fluxonium qubits. 
When biased at their half-flux quantum sweet spot, the chain is equivalent to the TFIM.  
TFIM has been used to encode optimization problems and is a common tool 
for quantum computing \cite{lucas2014ising, mcgeoch2014adiabatic}.
Further important insights into condensed matter applications of the fluxonium chain come from
the mapping of the system onto a chain of fermions using the standard Jordan-Wigner
transformation \cite{suzuki2012quantum}. 
Mapping of the Kitaev chain onto the TFIM may help to study topologically protected quantum states or Majorana bound states \cite{kitaev2001unpaired}. Even though the Ising chain does not have topological protection, 
the possibility to explore Majoranas with the chain of fluxonium qubits may shed light
on the feasibility of the Majorana-based quantum computing~\cite{backens2017emulating}.

In the chain formed by fluxonium circuits, imperfections of the fabrication and flux-tuning will produce disorder.  For example, the fabrication of the small Josephson junction of the fluxonium results in a variation of the Josephson energy and, consequently, in the qubit energy splitting.  
We evaluate conditions for the Josephson energy fluctuations to be weak enough for the chain excitations to remain delocalized. Then, we investigate how random flux detuning from the sweet spots brings the system to the localized regime. We demonstrate that
random magnetic fluxes through different fluxoniums can realize quenched disorder, allowing one to experimentally study statistical properties of many disorder realizations using a single device.

\emph{Fluxonium chain as transverse field
Ising model --}
The fluxonium qubit consists of a phase-slip Josephson junction shunted by a large inductor, commonly formed by a long Josephson junction array with the total inductance $L$.  The Hamiltonian for the fluxonium 
has the form \cite{manucharyan2009fluxonium}:
\begin{equation}
H_{l} = 4 \, E_C  \, n_l^2 + V(\theta_l), \quad V(\theta_l)=\frac{E_L }{2}  \, \, \theta_l^2 - E_{J,l} \, \cos(\theta_l - \phi_{l}),
\label{eq:Hamiltonian_Fluxonium}
\end{equation}
where $n_l$ and $\theta_l$ are the Cooper pair number
and phase operators, respectively. 
These operators satisfy 
the following commutation relation $[\theta_l,n_l] = i$.
Hamiltonian~\eqref{eq:Hamiltonian_Fluxonium} is characterized by three energies: the charging energy $E_C = e^2/2C$,  the Josephson energy of the phase-slip junction $E_J$, and  the inductive energy $E_L = \Phi_0^2/(4\pi^2 L)$, where
$\Phi_0=h/2e$ is the flux quantum, $h=2\pi \hbar$ is the Planck constant.
In the presence of the external magnetic flux
$\Phi_{l}$, the energy of the junction is offset by 
$\phi_{l} = 2\pi \Phi_{l}/\Phi_0$.

We assume that all fluxoniums are flux biased near the half-flux quantum sweet spots, $\phi_{l} =  \pi$, where the fluxonium exhibits its longest coherence times \cite{nguyen2019high}.
We consider a high-frequency fluxonium 
with $E_C/h=1.45$ GHz, $E_L/h=4.0$ GHz, and $E_J/h=9.0$ GHz.  
The energy splitting for 0-1 transition $\Delta E_{0-1} /h \simeq 2.0 $ GHz is significantly lower than the splitting between states 1 and 2, $\Delta E_{1-2}/h\simeq 10.2$ GHz.  For further discussion, we  restrict our analysis of the fluxonium dynamics by taking into account only its two lowest energy states. 

The interaction between fluxoniums is realized via a few shared junctions of superinductor arrays. The energy of interaction is determined by the common phase drop along shared junctions and is proportional to the product of the phase operators for the pair of fluxoniums
\cite{kou2017fluxonium}:
\begin{equation}
H_{\rm int} = \sum_{l=0}^{L-2} J_{l} \theta_l \theta_{l+1}, 
\quad
J_{l} = \left(\frac{\hbar}{2e}
\right)^2\frac{M_{l}}{L_lL_{l+1}}.
\end{equation}
Here, $L_l$ is the effective inductance of fluxonium $l$ and $M_{l}$ is the mutual inductance of neighboring fluxoniums $l$ and $l+1$.  Depending on the fraction of the shared Josephson junctions, the interaction between fluxoniums can reach relatively large values, $J\simeq E_L$.

In the eigenstate basis of individual fluxoniums, the Hamiltonian of the system becomes 
\begin{equation}
H_{\rm chain} = - \, \frac{1}{2} \sum_{l=0}^{L-1}  \varepsilon_l \hat \sigma^{z}_{l}+ J \sum_{l=1}^{L-2} \hat \theta_{l} \hat \theta_{l+1}
,
\label{eq:TFIM_detuned}
\end{equation}
where $\varepsilon_l=\varepsilon_l(\delta\phi_l)$ is the flux-dependent level spacing between the ground and first excited qubit states, $\hat \sigma^{\alpha}_l$ is the $\alpha=x,y,z$ Pauli matrix, and the phase operators are represented by the  matrices:
\begin{equation}
\hat \theta_l = \left(
\begin{array}{cc}
\theta_{gg}(\phi_l) & \theta_{ge} (\phi_l) \\
\theta_{eg} (\phi_l) & \theta_{ee} (\phi_l)
\end{array}
\right).
\label{eq:theta}
\end{equation}
For fluxonium at the half-flux quantum sweet spot, we have $\hat \theta_l (\phi_l=\pi)= a \hat \sigma_z$, where $a\approx 2.36 $ for the chosen fluxonium parameters. 
Below, for the interaction strength between fluxoniums, we take $J_{l}=J = 20$MHz.

We consider the chain dynamics when all fluxoniums but one are initialized in their ground states, and the only fluxonium $l=0$ is in its first excited state.  
This excitation can move to its neighbor in time $\propto \hbar/J$.  
This single excitation of the chain will move to other fluxoniums until it reaches the opposite end of the chain, then gets reflected and moves back.  
Solving the Schr\"odinger equation for the state of chain of fluxoniums, we calculate the excitation probability of fluxonium $l$ as:
\begin{equation}
p_l(t) = \bra{\psi(t)} \hat P_l \ket{\psi(t)}\, ,  
\end{equation}
where $\hat P_l =\ket{e_l}\bra{e_l}$ is the projection operator to the excited state of qubit $l$.

We present the evolution for a chain of $L=17$ identical fluxoniums in Fig.~\ref{fig:1}(b).    We estimate that the time of excitation propagation through the chain is given by $(M-1)\hbar/J=12$ ns.  We also note that the maxima of $p_l(t)$ are lower for the qubits with a higher index $l$ as the dispersion of the propagating excitation waves leads to the broadening of $p_l(t)$ and lowering its maxima.  The speed of excitation propagation is consistent with that of the TFIM~\cite{suzuki2012quantum}. The maximal probability increases closer to the ends of the chain as the reflection causes refocusing of the excitation wave packet there, compare $l=0$ and $l=16$ with other intermediate qubit locations.

\begin{figure}
    \centering
    \includegraphics[width =0.97\columnwidth]{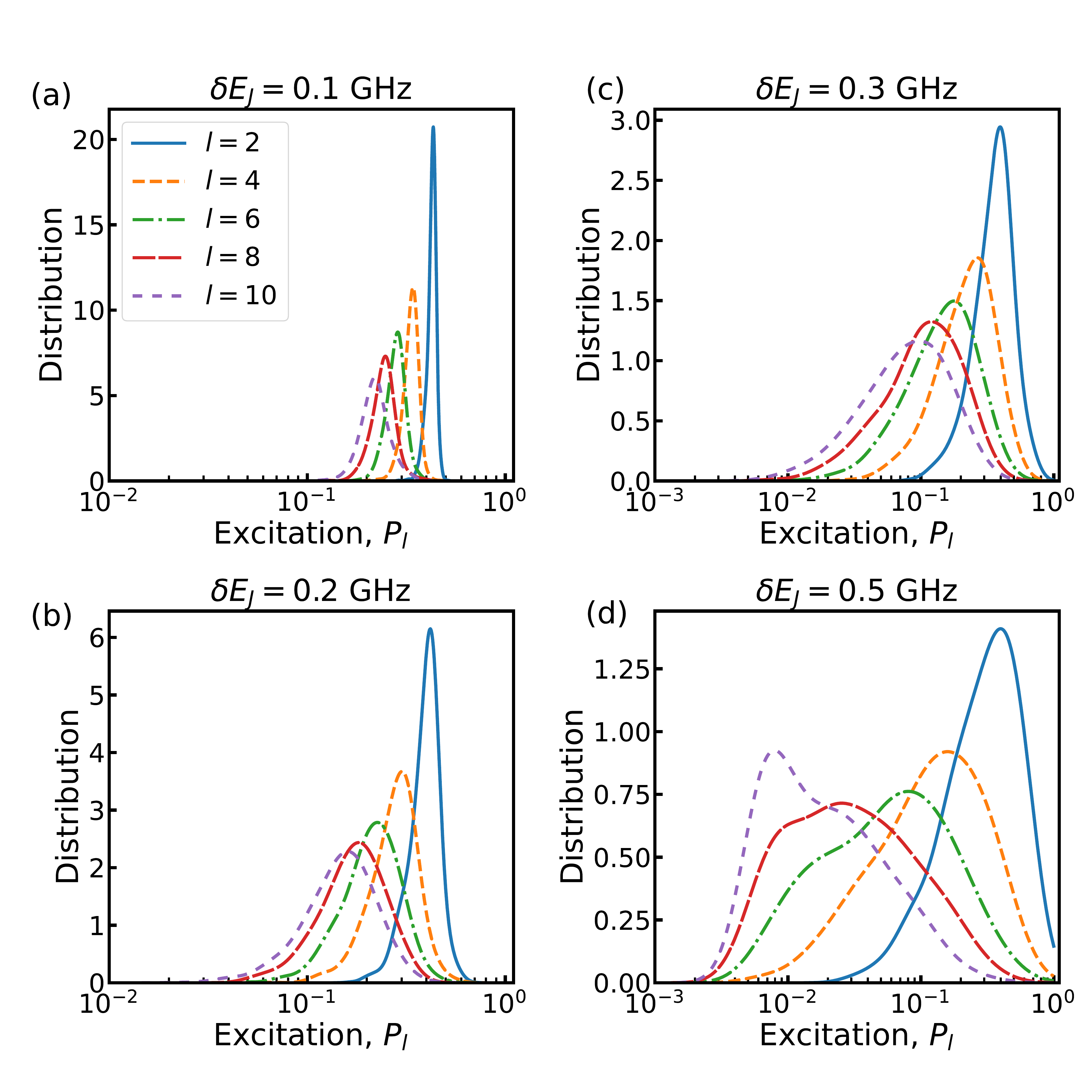}
    \caption{The distribution function of the excitation probability of
    spins $l=2$, 4, 6, 8, 10 in the chain with fluctuating Josephson energy $E_J$ with $\delta E_J=0.1$ GHz (a), 0.2 GHz (b), 0.3 GHz (c),  and $0.5$ GHz (d). At least 1000 disorder realizations are taken for all distributions in the paper.}
    \label{fig:3}
\end{figure}

\emph{Uncorrelated disorder --}  
The disorder in the fluxonium chain could originate from the fluctuations of $E_J$, mutual inductances $M_l$, and magnetic flux fluctuations $\phi_l$ through the fluxonium loop. 
The first two sources of the disorder are not easily adjustable for a specific device and are expected to be a consequence of fabrication imperfections. The magnetic field can be used as a synthetic disorder to study localization transition using the same device by randomly changing fluxes through the fluxoniums.

We assume that the interaction between qubits is uniform throughout the chain, $J_l = J$, which is determined by the number of shared Josephson junctions of the superinductor and is not expected to exhibit large fluctuations between fluxoniums of the chain.
On the other hand, the Josephson energy fluctuations $E_J$ is the major challenge to produce a uniform chain of fluxoniums.  Therefore, we
consider the effect of Gaussian fluctuations of $E_J$ on the chain dynamics and establish the allowance for its standard deviation $\delta E_J$ to have a chain in the delocalized regime. 
Thus, we assume that the Josephson energies of individual fluxoniums $E_{J,l}$  have a  Gaussian distribution with average value $\bar E_J$ and the standard deviation $\delta E_J$.  The main effect of fluctuations of $E_{J,l}$ is the random variation of energies $\varepsilon_l$ in Eq.~\eqref{eq:TFIM_detuned} with the standard deviation:
\begin{equation}
\label{eq:deltae}
\delta\varepsilon =\sqrt{\overline{\varepsilon_l^2} - \bar\varepsilon_l^2}\, .
\end{equation} 
Here the bar denotes averaging over disorder configurations. 
For small $\delta E_J \ll E_J$, we can use a perturbation theory to find the relation 
$\delta\varepsilon = \eta \delta E_J$,
where $\eta = |\bra{e}\cos\theta\ket{e} - \bra{g}\cos\theta\ket{g}| \approx 0.31$. 

To characterize the mobility of
excitations,  we 
study statistics of 
the maximal probability of excitation 
\begin{equation}
\label{eq:Pmax}
P_l = \max_{t< t_*} \, \, p_l(t)\, .
\end{equation} 
for times $t<t_*$, where $t_*\approx 30.0$ ns is twice the arrival time of the  excitation to the last qubit in the clean system, see Fig.~\ref{fig:1}(b). The results for the distribution function are presented in Fig.~\ref{fig:3}. For the weak disorder ($\delta E_J =0.1$ GHz, Fig.~\ref{fig:3}(a)) the distribution of $P_l$ remains narrow and is centered close to the values of $P_l^{(0)}$ for the clean system.  As disorder increases ($\delta E_J =0.2$ GHz, Fig.~\ref{fig:3}(b)) the distribution broadens and shifts to smaller values. At even stronger disorder, ($\delta E_J =0.3$ GHz), Fig.~\ref{fig:3}(c)) the   distribution of $P_l$ shifts to even smaller values 
with more than half of the realizations showing $P_{l}<0.1$ for $l=8$ and $10$.  For $\delta E_J=0.5$ GHz, the excitation propagates to nearest  qubits only and is extremely unlikely to reach qubits further away, as illustrated by the broad distributions for $l\geq 6$ with tails going below $0.01$. We notice that the distributions for more distant qubits, i.e. larger $l$, is broader and shifted to smaller values of $P_l$.

\begin{figure}
    \centering
        \includegraphics[width =0.97\columnwidth]{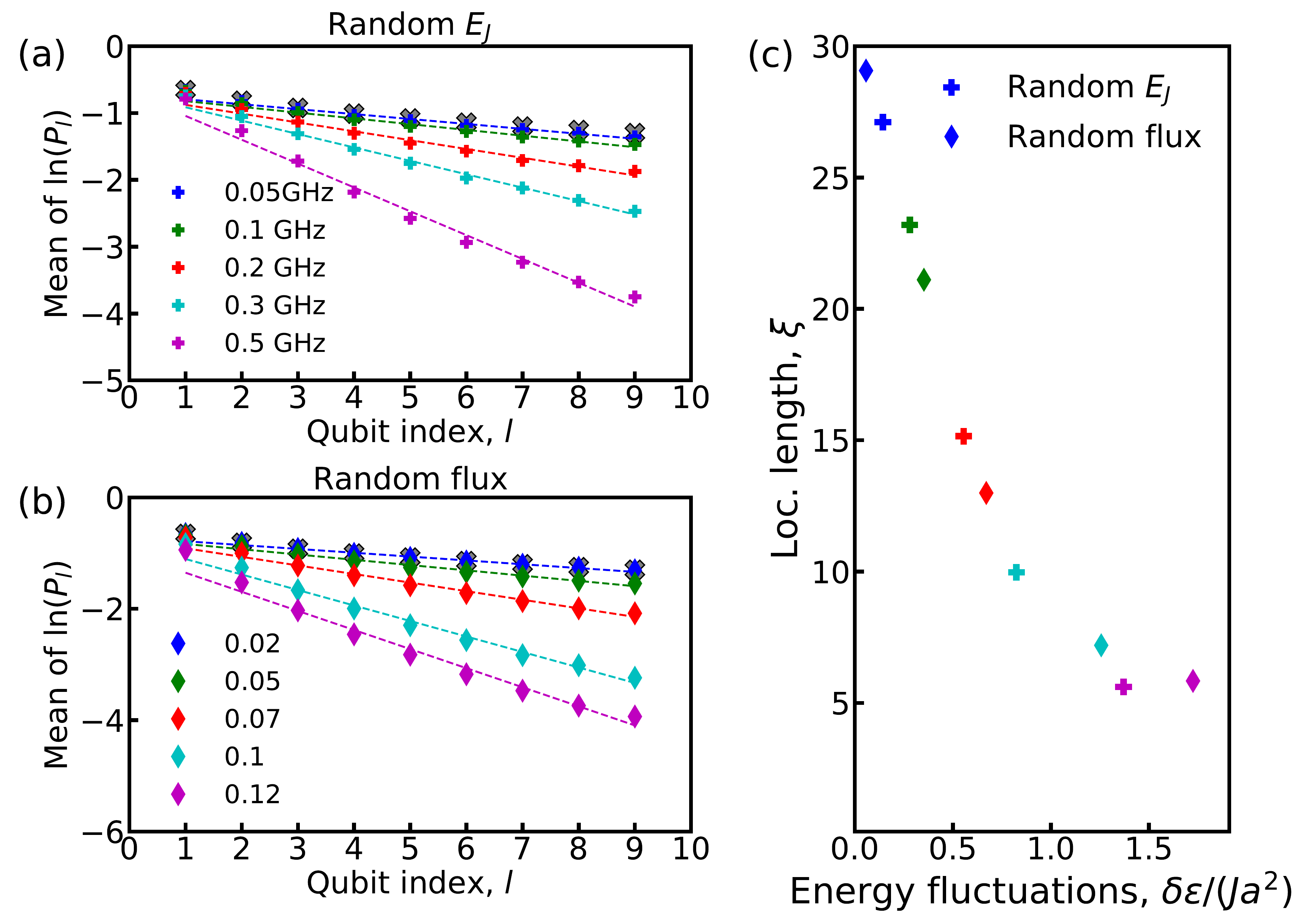}
    \caption{Mean value of the excitation probability as a function of spin index $l=1,\dots, 9$ with (a) fluctuating Josephson energy with standard deviation $\delta E_J=0.05\dots 0.5$ GHz and zero flux detuning from the sweet spot, and  (b) flux disorder with standard deviation $\delta\phi =0.02\dots 0.12$ and fixed $E_J/h=9$ GHz. The thick x-marks show the maximal excitation probability for the uniform chain when all qubits are identical and biased at the sweet spot.  Panel (c): the localization length, obtained from the linear fit of data in panels (a) and (b), shown as functions of the energy fluctuations $\delta\varepsilon$, Eq.~\eqref{eq:deltae}.}
    \label{fig:4}
\end{figure}

\emph{Localization length --} 
To quantify the suppression of excitation propagation through the chain, we evaluate the average value of  $\ln P_l$ for different qubit locations $l$, see Fig.~\ref{fig:4}(a).  
We observe that 
$\overline{\ln P_l} \approx c_0 - 2l/\xi$, where $\xi$ has a meaning of the localization length that characterizes decay of wave functions as $\exp(-x/\xi)$.  To avoid the effect of reflection at the opposite end of the chain, we perform the linear fit for $l\leq L/2$ to obtain values of $\xi$, as illustrated by dashed lines in Fig.~\ref{fig:4}(a).  We presented $\xi$ for several values of $\delta E_J$ in the range from $0.05$ to $0.5$ GHz in Fig.~\ref{fig:4}(c).  The horizontal axis shows the value of the qubit energy fluctuations 
$\delta\varepsilon$, Eq.~\eqref{eq:deltae},
calculated for corresponding values of $\delta E_J$ and made dimensionless by dividing by $Ja^2$.  
We notice that for weak disorder $\delta E_J \lesssim 0.2$ GHz, $\xi $ is comparable to the system size $L=17$.  In this limit, $\overline{\ln P_l}$ are approaching the values obtained for a uniform chain of fluxoniums at the sweet spot, when reduction of $P_l^{(0)}$ (marked by thick "x") with increasing $l$ is due to dispersion broadening of the excitation wave packet.
At stronger disorder, the localization length saturates to the values $\xi\simeq 1$, when only partial excitation propagation occurs to nearest neighboring sites. 
We identify the localization transition when $\xi \simeq L/2$ that occurs at $\delta E_J \simeq 0.3$ GHz or $\delta \varepsilon/(Ja^2)\simeq 0.5$.

This estimate for  the onset of the localization is consistent with the behavior of the distribution functions, presented in Fig.~\ref{fig:3}.  In weak disorder, the probability density vanishes for small $P_l\lesssim 0.1$, Fig.~\ref{fig:3}(a,b).  At $\delta E_J =0.3$ GHz shown in Fig.~\ref{fig:3}(c), the distribution  moves to the left and spreads over two orders of magnitude,  with non-zero probability to have  $0.01 < P_l< 1$. As disorder increases more, $\Delta E_J =  0.5$ GHz, values of $P_l<0.01$ become common, see Fig.~\ref{fig:3}(d).

\emph{Excitation at the edges --} We discussed the excitation dynamics of a qubit at an arbitrary location along the chain.  The excitation probability  of each qubit can be measured. However, such measurements are hard in near-term quantum simulators. 
Readout of individual qubits is usually slow, while a typical excitation time of the qubit is characterized by $h/(2Ja^2)$ and is well under 10 ns long in our case.  
Below we focus on the excitation dynamics of the last qubit, $l=16$.  
An edge qubit is easier to be coupled to a readout resonator without disturbing its neighboring qubits, and its excitation can be locked in time by applying a fast-flux detuning to its neighbor.  In this case, the large energy mismatch for the two qubits forbids the energy exchange between neighboring fluxoniums~\footnote{Numerical simulations show that if one qubit is tuned to the integer flux quantum sweet spot, the probability of the excitation for the other qubit stays intact.}. In such flux configuration, slow readout of the fast propagating qubit excitation is possible.  

\begin{figure}
\centering
\includegraphics[width =0.97\columnwidth]
{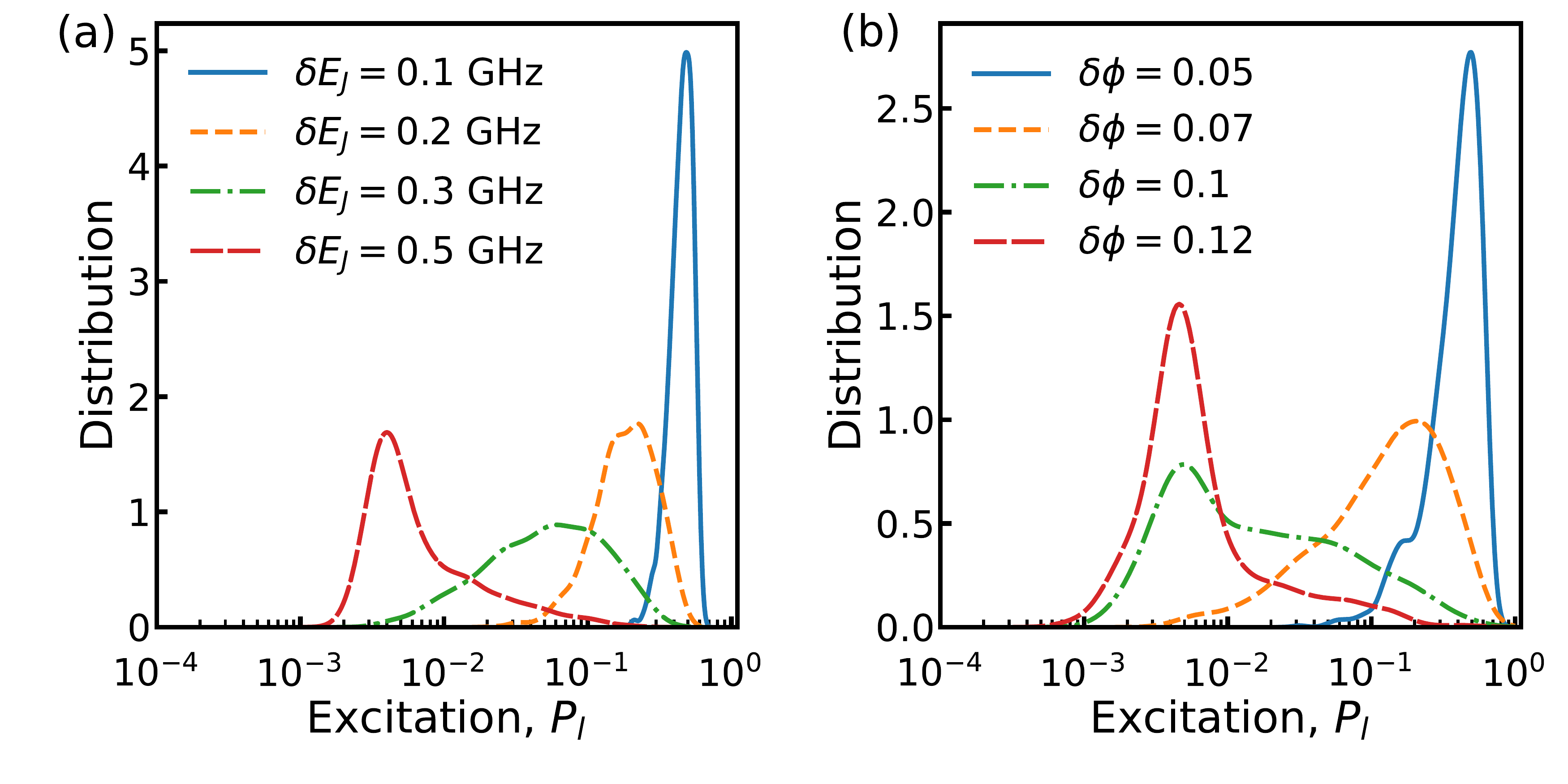}
\caption{The distribution function of the logarithm of the excitation probability $P_{l=16}$ of
spin $l=16$.
Panel (a): All fluxoniums are at the sweet spot and the distribution functions represent different strengths of fluctuations of the Josephson energy, $\delta E_J=0.1$, 0.2, 0.3, and 0.5 GHz. Panel (b): All fluxoniums have the same Josephson energy $E_J/h=9$ GHz with random flux detunings from the sweet spot with standard deviation $\delta\phi =0.05$, 0.07, 0.1, and 0.12. 
}
    \label{fig:5}
\end{figure}

The remote qubit maximal excitation probability  $P_{16}$ characterizes the propagation of excitations through the chain.  In weakly  disordered chain, $\delta E_J=0.1$ GHz, the maximum of excitation probability $P_{16}$ is a narrow distribution over disorder ensemble and is slightly reduced from $P_{16}^{(0)}\approx 0.67$, when compared to the clean system.  The distribution function vanishes for $P_{16}<0.1$, see Fig.~\ref{fig:5}(a). As the disorder strength increases, e.g. as $\delta E_J=0.2$ GHz, and the distribution of  $P_{16}$ shifts to smaller values and broadens. At $\delta E_J=0.3$ GHz which we identified above as the transition point to the localized regime, the distribution of  $P_{16}$ become the broadest with the range
$ 10^{-3} \lesssim P_{16}\lesssim 1$.  At larger fluctuations of the Josephson energy, e.g. $\delta E_J=0.5$ GHz, the distribution narrows and is centered at the smaller values of $P_{16}$, with the majority of $P_{16} < 0.01$.

\emph{Flux Disorder -- }
Here, we also consider a chain with random flux detunings 
from the sweet spot $\phi_l-\pi$.  We describe the localization onset as the standard deviation $\delta\phi$ of the flux detunings increases. We note that the flux detunings effectively introduce the longitudinal field, and thus the system acquires deviations from the TFIM (see the Appendix for more details). 

First, we evaluate $\overline{\ln P_{l}}$ for several values of flux disorder strength, characterized by $\delta \phi$.  We observe that $\overline{\ln P_{l}}$ is well-fitted by the linear dependence on the qubit location $l$ and the fitting coefficient provide the localization length in the chain, see Fig.~\ref{fig:4}(b).  We plot the localization length as a function of the dimensionless qubit energy fluctuations, $\delta\varepsilon/(Ja^2)$, in  Fig.~\ref{fig:4}(c).  We notice that the localization lengths obtained for the chain with either fluctuating Josephson energy or flux align well along the same curve when the localization length is plotted as a function of the dimensionless energy fluctuations.  Thus, we expect that the dominant cause of single excitation localization in the chain is the energy mismatch between neighboring qubits and is consistent with the Anderson localization~\cite{anderson1958absence}.  

We also present statistics of $P_{16}$
for the excitation of the last qubit for four values of flux disorder in Fig.~\ref{fig:5}(b).  The behavior of the distribution functions is similar to the cases with fluctuations of the Josephson energy.  For weak flux noise, $\delta\phi\lesssim 0.07$, the distributions are narrow, and the excitation probabilities mostly exceed $0.1$.  At the transition, which happens around $\delta\phi=0.1$, the distribution becomes broad.  It shifts to much smaller values of $P_l$, and observing the remote spin's excitation is very unlikely at stronger flux disorder. We again observe a qualitative similarity between disorder introduced by fluctuations of the Josephson energy and random magnetic flux detunings.

\emph{Conclusions -- } The chain of 
fluxonium qubits with record-high coherence 
provides means for
studying the effects of single-
and many-body localization of excitations 
in the coherent setting, as well as
multiqubit tunneling and multiqubit gates
due to strong inductive coupling. 
We studied 
fluxonium qubit systems simulating clean and
disordered TFIM.  
The spin-flip experiment identifies ergodic and localized regimes in the TFIM that require measurements of edge qubits.  Such measurements can be performed for a simple fluxonium chain and do not require a complicated circuit design.
Our results show that the fluxonium chain mimics TFIM well
and is a prominent candidate to be a near-term quantum
simulator of strongly correlated spin systems.  While we analyzed the properties of one-dimensional chain in this paper, fluxonium devices can also be used to explore two-dimensional lattices and Cayley trees.

The ideal starting point to study spin systems is to fabricate a uniform chain of fluxonium qubits and generate random flux detuning from the sweet spot to study the effects of disorder in the TFIM. However, the current fabrication process often provides about 10\% fluctuations of the Josephson energy.  These fluctuations are sufficient to bring the chain to the localized regime at the coupling strength considered in this paper.  The fabrication process has to be improved to reduce fluctuations of the Josephson energy of the fluxonium qubit and to achieve a delocalized regime.  
Alternatively, the interaction between qubits can be made stronger, thus bringing the system to the delocalized regime even if fluctuations of $E_J$ remain significant.
For fluxoniums with parameters analyzed here, the coupling between them has to be about $60$ MHz, which is still weak compared to the achievable galvanic coupling between fluxoniums~\cite{kou2017fluxonium}.

\emph{Acknowledgments -- }
We thank  Mark Dykman, and Kostya Nesterov for fruitful discussions. This work was supported by the U.S. Department of Energy, Office of Science, Office of Basic Energy Sciences, under Award Number DE-SC0019449. The work of A.B.\"O.  at Fermilab was supported by the DOE/HEP QuantISED program grant Large Scale Simulations of Quantum Systems on HPC with Analytics for HEP Algorithms (0000246788). This manuscript has been authored by Fermi Research Alliance, LLC under Contract No. DE-AC02-07CH11359 with the U.S. Department of Energy, Office of Science, Office of High Energy Physics. The simulations were performed using QuTiP \cite{johansson2013qutip} and the computing resources of the UW-Madison Center For High Throughput Computing (CHTC) and resources provided by the Open Science Grid \cite{pordes2007open, sfiligoi2009pilot}, which is supported by the National Science Foundation award 1148698 and the U.S. Department of Energy's Office of Science.

\appendix*

\begin{figure}[!htbp]
\begin{centering}
\includegraphics[width=0.98\columnwidth]{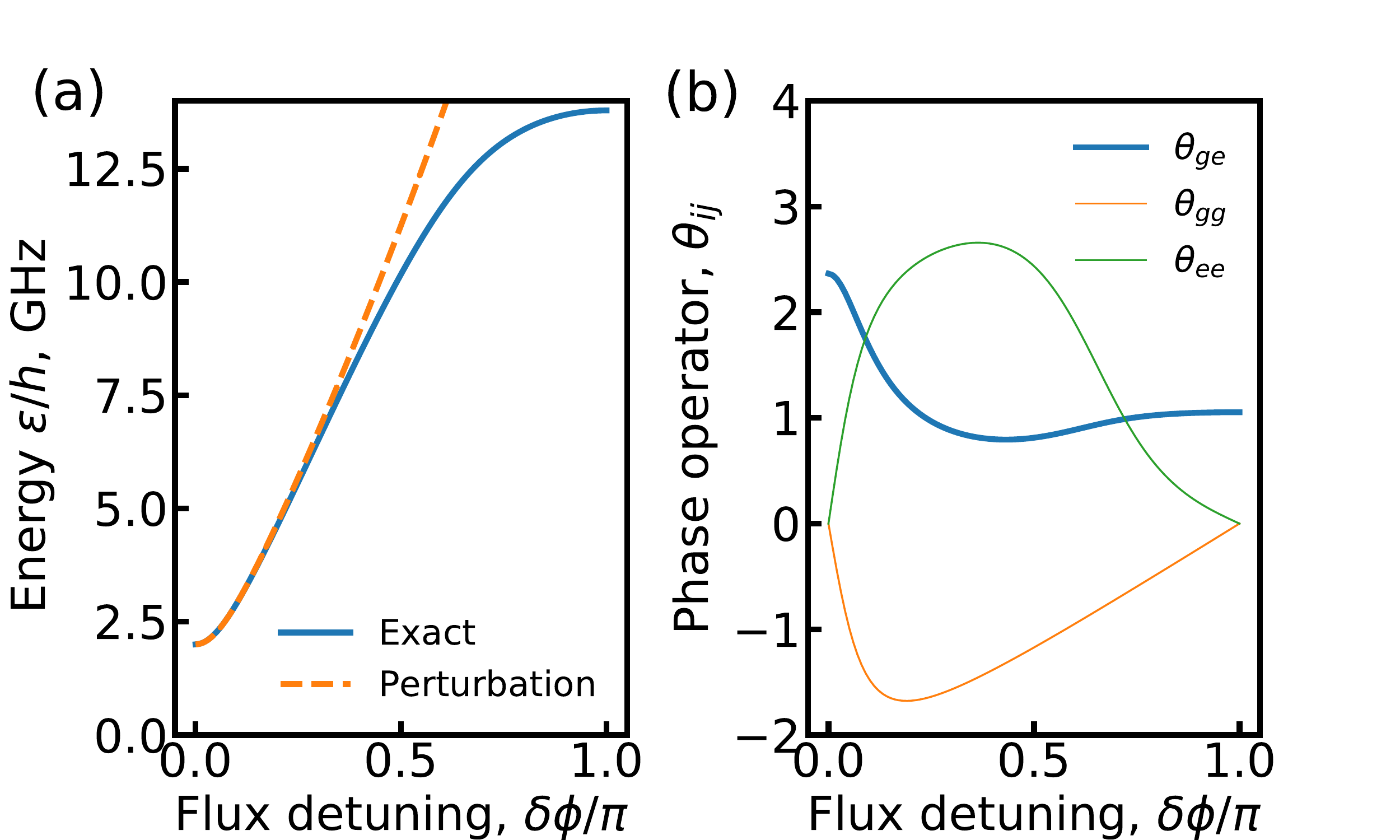}
\par\end{centering}
\caption{\label{fig:FluxoniumCircuit_and_EnergyDiagram}
(a) Dependence of the lowest excitation energy of a fluxonium as a function of the magnetic flux detuning from the half-integer sweet spot.
(b). Matrix elements for the phase operator $\hat \theta_l(\delta\phi)$ defined by Eq.~\eqref{eq:theta} are shown as a function of the magnetic flux detuning $\delta\phi$.  The excitation matrix element of the phase operator, $\theta_{ge}$, has maximum at the sweet spot and then decreases as detuning increases.  The diagonal matrix elements, $\theta_{gg}$ and $\theta_{ee}$, vanish at the sweet spots $\phi=\pi$ and $\phi=0$ but are finite when qubits are detuned from their sweet spots.
}
\label{fig:A1}
\end{figure}

\section{Fluxonium chain away from the sweet spot}

The Hamiltonian of the chain is given by Eqs.~\eqref{eq:TFIM_detuned} and \eqref{eq:theta}.  In this Appendix, we focus on the system of identical fluxoniums but consider small detuning from the half-flux quantum sweet spot with a random distribution of $\delta \phi_l$. We analyze how detuning from the half-integer flux quantum sweet spot changes the energy and phase matrix elements.  We also analyze how the uniform flux detuning affects the dynamics of excitation propagation through the chain.  
We present the plot for the dependence of the qubit energy splitting as a function of the flux detuning $\delta \phi_l = \phi_l-\pi$ 
in Fig.~\ref{fig:A1}(a).  We notice that the minimal energy splitting is at the half-integer sweet spot and then monotonically increases to the maximal splitting at the integer sweet spot.  In Fig.~\ref{fig:A1}(b), we present
the dependence of matrix elements of the phase operator $\hat \theta$ on the flux detuning, written in the basis of the eigenstates of the fluxonium Hamiltonian.  The off-diagonal matrix element, $\theta_{eg}$ responsible for the $XX$ coupling at the sweet spot decreases from its value at the sweet spot, where $\theta_{eg}(\phi= \pi) =a\approx 2.36$. 
The off-diagonal matrix elements $\theta_{gg}$ and $\theta_{ee}$ vanish at the sweet spot, but their magnitude increases fast with detuning, see Fig.~\ref{fig:A1}(b).

To better illustrate the properties of the fluxoniums away from the sweet spot, we also analyze the system Hamiltonian in the basis of fluxonium eigenstates at the sweet spot.  In this basis, the phase operator $\theta = a\sigma_x$ with $a\approx 2.36$ for small flux detuning and the choice of parameters introduced in the main text.  The Hamiltonian of the system in this basis and at small detuning has the form 
\begin{equation}
H_{\rm TFIM} = 
Ja^2 \sum_{l=0}^{L-2} \hat \sigma^x_l \hat \sigma^x_{l+1}
+  \sum_{l=0}^{L-1} H_{l}
,
\label{eq:TFIM_Fluxonium}
\end{equation}
where 
the Hamiltonian of qubit $l$ in the basis of eigenstates of the fluxonium at the sweet spot is
\begin{equation}
H_{l} = -\frac{\varepsilon_0+\delta \varepsilon^z_l}{2} \hat \sigma^z_l -\frac{\delta \varepsilon^x_l}{2} \hat \sigma^x_l , 
\label{eq:H_qubit}
\end{equation}
where $\varepsilon_0 =(E_1-E_0)$ is given by the difference of the qubit energies at the sweet spot.

We now evaluate fields  $\delta\varepsilon^{z}$  and  $\delta\varepsilon^{x}$ for small deviations of the flux $\delta\phi_l$.
The phase-dependent energy $V(\theta)$ acquires the correction
$\delta V(\theta) =  E_J (\delta\phi \sin\theta -(\delta\phi^2/2) \cos\theta) $.  We apply the perturbation theory in the basis of eigenstates of the fluxonium at the sweet spot. We find that 
flux-dependent energy shift determined by the diagonal matrix elements of $\cos\theta$, and $\delta \varepsilon^x$ is determined by the off-diagonal matrix element of $\sin\theta$:
\begin{eqnarray}
\delta \varepsilon^z_l & = & \frac{E_J (\delta\phi_l)^2}{2}[\bra{e}\cos\theta \ket{e} -\bra{g}\cos\theta\ket{g} ],\\
\delta \varepsilon^x_l & = & 2E_J\delta \phi_l \bra{g}\sin\theta\ket{e}.
\end{eqnarray} 
This Hamiltonian corresponds to the transverse Ising Hamiltonian
with random transverse and longitudinal fields. We note that the contribution of $\delta \varepsilon_l^z$ is quadratic in the flux detuning and, therefore, the main contribution of the fluctuating field is equivalent to a random longitudinal field.  When Hamiltonian~\eqref{eq:TFIM_Fluxonium} is rewritten in the eigenstate basis of the individual qubits, it acquires the form given by Eq.~\eqref{eq:TFIM_detuned}, when the interaction is no longer represented by $\hat\sigma^x_l\hat\sigma^x_{l+1}$ terms only, at the same time, the coefficient in front of   $\hat\sigma^x_l\hat\sigma^x_{l+1}$ term is reduced as the phase matrix element $\theta_{eg}$ decreases, see Fig.~\ref{fig:A1}(b).

\begin{figure}[!htbp]
    \centering
    \includegraphics[width =0.9\columnwidth]{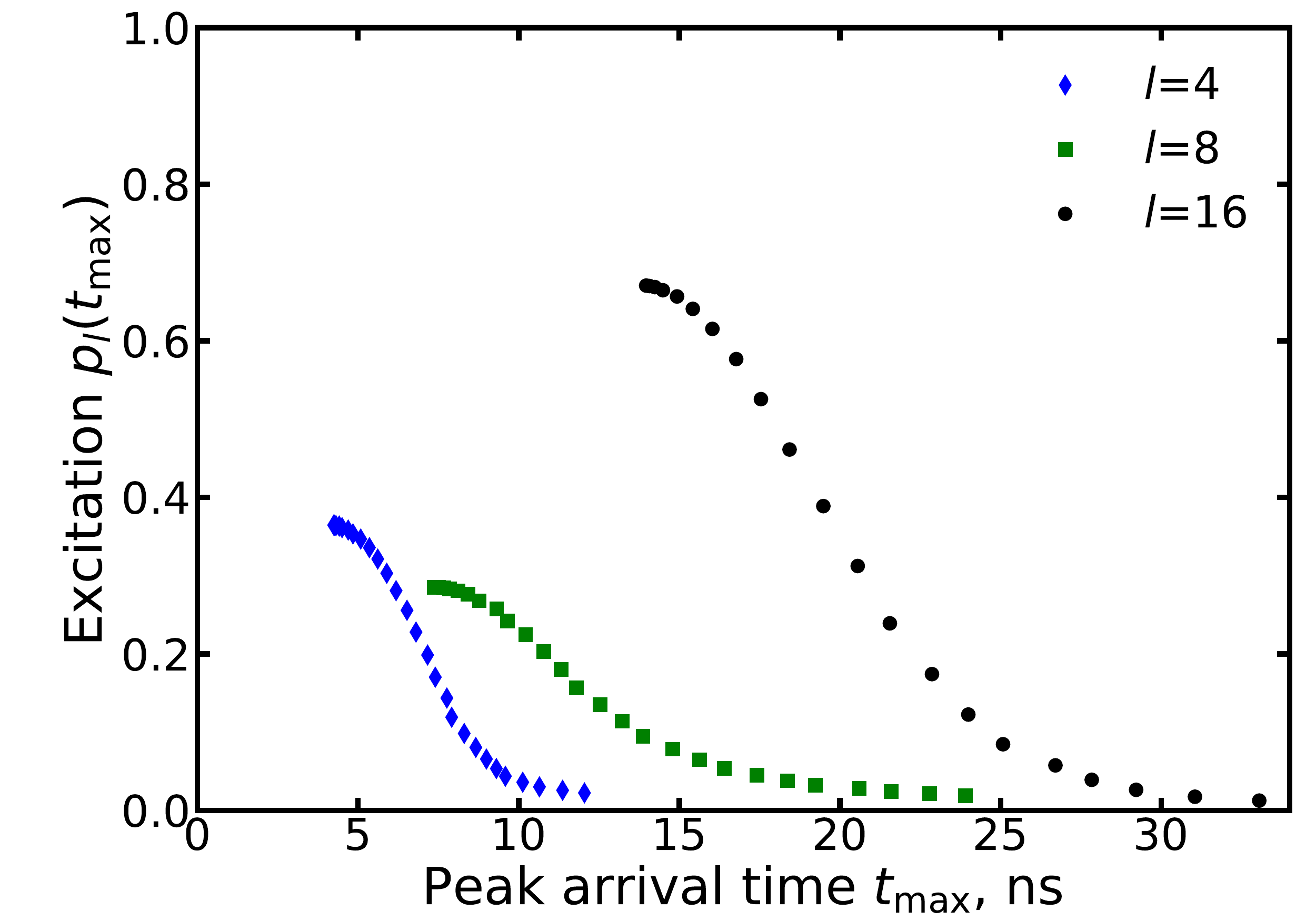}
    \caption{The parametric plot of the peak excitation probability for  
    qubits $l=4,8,16$ as a function of the peak arrival time $t_{\rm max}$ in the homogeneous system of $L=17$ fluxoniums with flux detuning from the sweet spot. Initially, all qubits were in their ground states, but fluxonium $l=0$ was excited.  The parameters of fluxoniums and the coupling strength are provided in the text, $J=10$ MHz.
    }
    \label{fig:A}
\end{figure}

In the main text, we discussed the chain dynamics with all qubits initialized in their ground states and only qubit  $l=0$ in its first excited state.  The dynamics is characterized by the excitation probability of different qubits as a function of time.  We can evaluate the location of the peaks and their height for different qubits. The peak time is consistent with the maximal group velocity of the TFIM estimated from the spectrum of the TFIM~\cite{suzuki2012quantum}:
\begin{equation}
u(q)=\frac{\partial \omega_q}{\partial q}=\frac{1}{\hbar}\frac{J a^2 \varepsilon_0\sin q}{\sqrt{\varepsilon_0^2+2Ja^2 \varepsilon_0\cos q+J^2a^4}},
\end{equation}
where $q$ is the quasimomentum for excitation with energy $\hbar\omega_q$.  For $q=\pi/2$ and $\varepsilon_0\gg Ja^2$, we have $u(\pi/2) \approx Ja^2/\hbar$.

We now explore the excitation propagation when the flux detuning from the sweet spot is identical for all fluxoniums.  As we argued above, a weak detuning from the sweet spot is equivalent to the additional longitudinal field applied to the TFIM.  As the detuning increases, individual qubits' level spacing increases but does not affect the resonant transfer of excitations between qubits. The main consequence of the detuning is a modification of the interaction term, which in the qubit eigenstate basis corresponds to the appearance of   $\sigma_l^z \sigma_{l+1}^z$ channel, while the $\sigma^x_{l} \sigma^x_{l+1}$ coupling decreases.  This decrease results in a reduced velocity of the excitation propagation and also in stronger dispersion. As a result, the propagation of the excitation in the system uniformly detuned from the sweet spot results in the longer propagation of the excitation along the chain and lower probability excitation of remote qubits due to the broadening of the excitation wave packet. We compute the relation between the peak excitation probability of qubits and the corresponding peak time.  The plot of the maximal excitation probability as a function of its arrival time is shown in  Fig.~\ref{fig:A}.  The monotonic decrease of the maximal excitation probability on the peak arrival time is consistent with a homogeneous reduction of the group velocity due to weaker excitation exchange in the longitudinal field and stronger dispersion.

\bibliography{MBL_Floquet}

\end{document}